\documentclass[preprint,12pt]{elsarticle}

\usepackage{graphicx}

\usepackage{amssymb}

\journal{Discrete Optimization}

\begin{document}

\begin{frontmatter}



\title{Approximations of the Densest k-Subhypergraph and Set Union Knapsack Problems}


\author[label1]{Richard Taylor}
\address[label1]{Defence Science and Technology Group, Canberra, ACT 2600, Australia}

\begin{abstract}
For any given $\epsilon>0$ we provide an algorithm for the Densest $k$-Subhypergraph Problem with an approximation ratio of at most $O(n^{\theta_m+2\epsilon})$ for $\theta_m=\frac{1}{2}m-\frac{1}{2}-\frac{1}{2m}$ and run time at most $O(n^{m-2+1/\epsilon})$, where the hyperedges have at most $m$ vertices. We use this result to give an algorithm for the Set Union Knapsack Problem with an approximation ratio of at most $O(n^{\alpha_m+\epsilon})$ for $\alpha_m=\frac{2}{3}[m-1-\frac{2m-2}{m^2+m-1}]$ and run time at most $O(n^{5(m-2)+9/\epsilon})$, where the subsets have at most $m$ elements. The author is not aware of any previous results on the approximation of either of these two problems.
\end{abstract}


\end{frontmatter}


\section{Introduction}
\label{intro}
In the Densest k-Subhypergraph Problem(DkSHP) we are given a hypergraph and an integer $k$, and wish determine the set of $k$ vertices such that the subhypergraph induced by this set has a maximum number of hyper-edges. This problem is a natural generalization of the Densest k-subgraph Problem(DkSP) and is in turn generalized to a programming problem by giving costs to the vertices and profits to the hyper-edges. Thus the Set Union Knapsack Problem(SUKP) \cite{GOL1994} has a set $S$ of $n$ items with a cost $c_i$ associated with each item $i$; a collection of subsets $P$ of $S$ each with a profit $p_e$ associated with each $e \in P$; and a cost bound $B$. $B$ as well as all costs and profits are non negative reals. The problem is to find a subset $U$ of $S$ whose total cost is bounded by $B$ with total profit maximized.
In mathematical terms this can be stated as
\begin{eqnarray}
\max_{U \subseteq S} \left\{ \sum_{e \in P, e \subseteq U} p_e \right\} \nonumber \\
\textrm{ subject to }  \sum_{i \in U} c_i \leq B.
\end{eqnarray}
The Densest k-Subhypergraph and Set Union Knapsack problems are generalizations of a number of other studied combinatorial problems. In Figure 1 we illustrate these generalization relationships \cite{TAY12015} with a link indicating the higher problem as a generalization of the linked problem below it. In some cases a problem can be solved by restrictions on another problem together with the need to apply the second problem a small number of times (polynomial in the input length).  If this is the case we indicate this with an asterix next to the original problem in the figure.

In this paper we provide an appoximation to the Densest k-Subhypergraph Problem, and together with a generalization of the method we gave in \cite{TAY22015} use this result to give an appoximation to the Set Union Knapsack Problem. Note that both of these results are in terms of an integer parameter $m$. In the Densest k-Subhypergraph Problem $m$ is the maximum number of vertices in any hyperedge, while in the Set Union Knapsack Problem $m$ is the maximum number of vertices in any subset in $P$.
\begin{figure}
\begin{center}
\includegraphics[width=100mm]{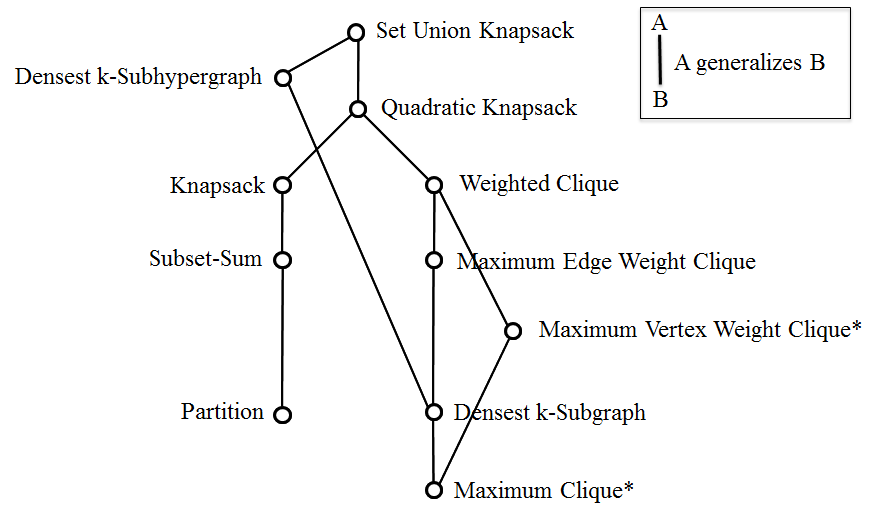}
\caption{\label{figure1.png} Set Union Knapsack problem and tree of sub-problems}
\end{center}
\end{figure}
\section{Approximating the Densest k-Subhypergraph problem}
Let $G$ be any hypergraph of order $m$ where all the hyper-edges consist of vertex subsets of size $m$. \\
Notation: Let an $m$-hyperedge be a hyperedge involving $m$ vertices. Let $E^m[G]$ be the collection of $m$-hyperedges of $G$. Let $MAX^m[G,k]$ be the maximum number of $m$-hyperedges in any selection of $k$ vertices of $G$, and $G^{MAX}[k]$ be a subhypergraph of $G$ with $MAX^m[G,k]$ $m$-hyperedges. 

We provide three lemmas that are used in the main result of this section. \\ \\
\textbf{Lemma 2.1.} {\em There is a $O(k^{m-1})$ approximation algorithm for DkSHP.} \\ \\
\textbf{Proof.} If $G$ has less than $k/m$ $m$-hyperedges then if all $m$-hyperedges are selected they are incident with at most $k$ vertices and so we can solve DkSHP optimally. If $G$ has greater than or equal to $k/m$ $m$-hyperedges then any selection of $k/m$ $m$-hyperedges must use at most $k$ vertices. Since any $k$ vertices can contain at most ${k \choose m} \leq k^m$ $m$-hyperedges and $k^m/(k/m)=O(k^{m-1})$ we have the result.\\ \\
\textbf{Lemma 2.2.} {\em $MAX^m[G,\lfloor k/2 \rfloor] \geq \frac{1}{3^m} MAX^m[G,k]$  for $k \geq 4m$.} \\ \\
\textbf{Proof.} Let $a_i, i=1,.., {k \choose \lfloor k/2 \rfloor}$ be the numbers of $m$-hyperedges in the subgraphs induced from subsets of $\lfloor k/2 \rfloor$ vertices chosen from $G^{MAX}[k]$. Then if we sum the 
$a_i$ this counts each $m$-hyperedge $k-m \choose \lfloor k/2 \rfloor-m$ times so that
\begin{equation}
\sum_{i}^{k \choose \lfloor k/2 \rfloor} a_i = {k-m \choose \lfloor k/2 \rfloor-m} MAX^m[G,k].
\end{equation}
It follows that for some $j$ 
\begin{eqnarray}
a_j &\geq& \frac{{k-m \choose \lfloor k/2 \rfloor-m}}{{k \choose \lfloor k/2 \rfloor}} MAX^m[G,k] \nonumber \\
&=& \left(\prod_{i=1}^{m} \frac{\lfloor k/2 \rfloor+1-i}{k+1-i}\right)MAX^m[G,k].
\end{eqnarray}
Since $k \geq 4m$ each term of the product is at least $1/3$ so it follows that
\begin{equation}
a_j \geq \frac{1}{3^m}MAX^m[G,k].
\end{equation}
Since $MAX^m[G,\lfloor k/2 \rfloor] \geq a_j$ the result follows. \\ \\
Notation: Let DkSHWP denote the Densest k-Subhypergraph Weighted Problem where the hyperedges have positive real weights and we seek the set of $k$ vertices such that the subhypergraph induced by this set has a maximum total hyperedge weight. In such a weighted hypergraph $G$ of order $m$ let $MAX^m[G,k]$ be the maximum edge weight induced by any selection of $k$ vertices of $G$. The following is essentially a result of \cite{FEI2001}.\\ \\
\textbf{Lemma 2.3.} {\em Let there be an algorithm for DkSHP ($m \geq 2$) that has an approximation ratio $O(f(n))$ and run time $O(g(n))$. Then there is an algorithm for DkSHWP ($m \geq 2$) that has an approximation ratio at most $O(f(n)log(n))$ and run time at most $O(g(n)log(n))$.} \\ \\
\textbf{Proof.} Let $G$ be any hypergraph of order $m$ where all the $m$-hyperedges have positive weights. Let $A$ be an approximation algorithm for DkSHP ($m \geq 2$) that has an approximation ratio of $O(f(n))$ and runs in time $O(g(n))$. Let $w$ be the maximum weight of any hyperedge in $G$. Obtain $G^*$ by rounding the hyperedge weights down to the nearest number among
\begin{equation}
w,\frac{w}{2},\frac{w}{2^2},...,\frac{w}{2^l},0, \textrm{  where  } l=\lceil log_2(x) \rceil, x={n \choose m}.
\end{equation}
Then
\begin{equation}
MAX^m[G^*,k] \geq \frac{1}{4}MAX^m[G,k]
\end{equation}
since the rounding down to non-zero edge weights and reducing those edge weights less than $w/{2^l}$ to $0$ each reduce the maximum total profit by a factor of at most $1/2$. Now use algorithm $A$ on each of $l+2$ problems, one for each subgraph of $G^*$ corresponding to each weight value (with all other hyperedges removed). Select the solution subset with the maximum weight value. This must produce a subgraph of $G^*$ on $k$ vertices with edge weight at least
\begin{equation}
\frac{1}{l+2}MAX^m[G^*,k] \geq \frac{1}{4(l+2)}MAX^m[G,k].
\end{equation}
Since $l=O(log(n))$ the result follows. \\ \\
\textbf{Theorem 2.1.} {\em Let $A$ be any approximation algorithm for DkSP that has an approximation ratio $O(n^{\alpha+\epsilon}), 0<\alpha<1$ and run time $O(n^{1/\epsilon})$. Then there is an approximation algorithm $A'$ for DkSHP ($m \geq 2$) that has an approximation ratio at most $O(log^{m-2}(n)n^{\theta_m+\epsilon})$ for $\theta_m=\frac{1}{2}m-\frac{1}{2}+\frac{1}{m}(2\alpha-1)$ and run time at most $O(n^{m-2+1/\epsilon})$. In particular this approximation ratio is less than $O(n^{\theta_m+2\epsilon})$.} \\ \\
Since \cite{BHA2010} provides an approximation algorithm for DkSP with ratio $O(n^{\frac{1}{4}+\epsilon})$ and runs in time $O(n^{1/\epsilon})$ the following corollary is immediate. \\ \\
\textbf{Corollary 2.1.} {\em There is an algorithm for DkSHP that has an approximation ratio less than $O(n^{\theta_m+2\epsilon})$ for $\theta_m=\frac{1}{2}m-\frac{1}{2}-\frac{1}{2m}$ and run time less than $O(n^{m-2+1/\epsilon})$.} \\ \\
\textbf{Proof of Theorem 2.1.} Let $G$ be a hypergraph on $n$ vertices with $m$-hyperedges. In the following analysis $m$ is considered fixed while both $k$ and $n$ are variables. The proof is by induction on $m$. For $m=2$ the result follows from \cite{BHA2010}. Assume therefore that the result is true for $m=r-1, r \geq 3$. We now consider the case where $m=r$. We analyse two cases. \\
Case 1  $k<n^{\theta_{r} /(r-1)}$. The result follows for this case by Lemma 2.1 (choose $m=r$).\\
Case 2  $k \geq n^{\theta_{r} /(r-1)}$. Let $l=\lfloor k/2 \rfloor$. Partition the vertices of $G$ into $\lceil n/l \rceil$ sets $s_1, s_2,..,s_{\lceil n/l \rceil}$ of size $l$ or smaller (at most one smaller set may be required). For any vertex $v$ of $s_i$ let $E_v$ be the set of $(r-1)$-hyperedges formed by the $r-1$ vertices other than $v$ in each $r$-hyperedge containing $v$. Define $G_i$ to be the multi-hypergraph with $(r-1)$-hyperedges
\begin{equation}
E(G_i)=\bigcup_{v \in s_i} E_v.
\end{equation}
Each $r$-hyperedge of $G$ is counted $r$ times over the $G_i$, thus for some $j$ we must have
\begin{equation}
MAX^{r-1}[G_j,k/2] \geq \frac{rMAX^r[G,k/2]}{\lceil n/l \rceil}.
\end{equation}
Now each $G_i$ as a multi-hypergraph with $(r-1)$-hyperedges can also be considered as an edge weighted hypergraph, with the weights corresponding to the number of $(r-1)$-hyperedges between any $r-1$ vertices. By our induction hypothesis and Lemma 2.3 we have an algorithm that takes time $O(log(n)n^{r-3+1/\epsilon})$ and can find a subgraph $G_j^*$ of $G_j$ on at most $\lfloor k/2 \rfloor$ vertices with 
\begin{equation}
|E^{r-1}[G^*_j]| \geq \frac{MAX^{r-1}[G_j,\lfloor k/2 \rfloor]}{s}
\end{equation}
for $s=O(log(n)log^{r-3}(n)n^{\theta_{r-1}+\epsilon})=O(log^{r-2}(n)n^{\theta_{r-1}+\epsilon})$. Combining this with inequality (9) we have
\begin{equation}
|E^{r-1}[G^*_j]| \geq \frac{rMAX^r[G,k/2]}{s \lceil n/l \rceil}.
\end{equation}
Since we do not know $j$ we need to apply the algorithm to all $\lceil n/l \rceil$ of the $G_i$ to find the subgraph $G_j^*$ with a maximum number of $(r-1)$-hyperedges. Since $\lceil n/l \rceil \leq 2n/k+1$ this takes an amount of time at most
\begin{equation}
O((\frac{2n}{k}+1)log(n)n^{r-3+1/\epsilon})=O(\frac{2n}{k}log(n)n^{r-3+1/\epsilon}).
\end{equation}
The lower bound on $k$ defining this case ensures that $(2n/k)log(n)<n$ for large $n$ so that this time bound is less than
\begin{equation}
O(n^{r-2+1/\epsilon}).
\end{equation}
Now form the induced subgraph $G_j^{**}$ of $G$ on those vertices formed by the union of the vertices of  $G_j^*$ with $s_j$. This reconstructs the $r$-hyperedges from the order $(r-1)$-hyperedges of $G_j^*$ but may count each $r$-hyperedge up to $r$ times in the process (for each $(r-1)$-hyperedge of $G_j^*$). Combining this observation with inequality (11) $G_j^{**}$ is a subgraph of $G$ with
\begin{equation}
|E^r[G_j^{**}]| \geq \frac{|E^{r-1}[G_j^*]|}{r} \geq \frac{MAX^r[G,k/2]}{s \lceil n/l \rceil}.
\end{equation}
Since $s_j$ has at most $\lfloor k/2 \rfloor$ vertices and $G_j^*$ has at most $\lfloor k/2 \rfloor$ vertices it also follows that $G_j^{**}$ has at most $k$ vertices. By Lemma 2.2 and using $\lceil n/l \rceil \leq 2n/k+1$
\begin{equation}
|E^r[G_j^{**}]| \geq \frac{MAX^r[G,k]}{3^rs \lceil n/l \rceil}\geq \frac{MAX^r[G,k]}{3^rs (2n/k+1)}.
\end{equation}
Since $s=O(log^{r-2}(n)n^{\theta_{r-1}+\epsilon})$ and by the lower bound for $k$ defining this case $G_j^{**}$ provides an approximation of $MAX^r[G,k]$ of order
\begin{equation}
O(log^{r-2}(n)n^{1+\theta_{r-1}-\theta_r/(r-1)+\epsilon}).
\end{equation}
It is elementary to verify that
\begin{equation}
1+\theta_{r-1}-\theta_r/(r-1)=\theta_{r},
\end{equation}
so that $G_j^{**}$ provides an approximation of $MAX^r[G,k]$ of order
\begin{equation}
O(log^{r-2}(n)n^{\theta_r+\epsilon}).
\end{equation}
This completes the induction step for the approximation result. The induction step for the time element following from the time bound (12). Finally since 
\begin{equation}
log^{m-2}(n)=n^{\frac{(m-2)loglog(n)}{log(n)}}
\end{equation}
and the exponent $(m-2)loglog(n)/log(n)$ is less than $\epsilon$ for $n$ large we also have the approximation ratio bounded above by $O(n^{\theta_r+2\epsilon})$. \\ \\
\textbf{Remarks} Note the form of the exponent $\theta_m=\frac{1}{2}m-\frac{1}{2}+\frac{1}{m}(2\alpha-1)$ in Theorem 2.1. For large $m$ this expression is dominated by the $\frac{1}{2}m$ term which follows from the inductive construction in the theorem. The $\alpha$ term corresponding to the approximation bound for the DkSP problem is significant for small $m$ but less so as $m$ increases. It is also easy to see that the results of this section are also the same if the hypergraphs $G$ are assumed to have hyperedges of size at most $m$ (rather than $m$ exactly).
\section{Approximating the Set Union Knapsack Problem}
Let $G$ be any hypergraph of order $m$ where all the hyper-edges consist of vertex subsets of size $m$. Let there be costs associated with the vertices $S$ of $G$ and profits associated with both the vertices and hyperedges of $G$. Thus SUKP takes the form of determining a set of vertices with total cost at most $B$ with the total profit of the induced subhypergraph on $S$ maximised. We extend the approximation results of DkSHP to SUKP. In doing so we use the term profit rather than weight. The method follows much the same approach as we gave in \cite{TAY22015} for the Quadratic Knapsack Problem (SUKP for $m=2$), but in this case generalizing to any $m$, and also using the approximation results for the DkSHP given in Section 2. \\ \\
\textbf{Theorem 3.1.} {\em There is an algorithm for SUKP that has an approximation ratio at most $O(n^{\alpha_m+\epsilon})$ for $\alpha_m=\frac{2}{3}[m-1-\frac{2m-2}{m^2+m-1}]$ and run time at most $O(n^{5(m-2)+9/\epsilon})$.} \\ \\
{\em Overview} \\
We sketch the main flow of the proof method. We group the costs and profits of the original instance into $O(logn)$ many buckets. This allows us to split the original instance into $O((logn)^{m+1})$ sub-instances each with a simplified structure. One of these sub-instances has costs and profits on the vertices only, while each of the others have closely bounded costs, no vertex costs and  the same edge profits. In this way we group these sub-instances into three classes: the first is an instance of the classical Knapsack Problem; while for the other two classes the SUKP can be approximated provided we have an approximation for the DkHS problem. Finally by an averaging argument at least one of the sub-instances must have a profit at most a factor of $O((logn)^{m+1})$ of the original instance. \\ \\
\textbf{Proof of Theorem 3.1.} The proof is by induction on $m$. For $m=2$ we have an approximation ratio of $O(n^{\frac{2}{5}}+\epsilon)$ in time $O(n^{9/\epsilon})$ \cite{TAY22015}. We assume therefore that the theorem holds for all $m<r$ and show that it follows for $m=r$.\\ \\
Notation: Let $MAX^r[G,B]$ be the maximum total profit of the solution to SUK applied to the Hypergraph $G$ of order $r$ with cost limit $B$. Let $G^{MAX}[B]$ be a subhypergraph of $G$ corresponding to a maximum solution. Let any subhypergraph in which the total cost of the vertices is at most $B$ be termed \textit{feasible}. For any subset $A$ of vertices let $cost(A)$ be the sum of all costs of vertices in $A$. \\ \\
{\em Pruning, Rounding and Grouping} \\
Let $G$ be a hypergraph. We may prune all vertices with $c_i > B$ and all edges $e$ where $\sum_{i \in e}c_i>B$ since these cannot be part of a feasible solution. Also if $p_e=0$ then the edge $e$ can be pruned. Let $p^*$ be the largest $p_e$ and $2^l$ the largest power of $2$ at most $p^*$.  Now form the graph $G'$ from $G$ with each $p_e$ rounded down to the nearest profit among $\{2^l,2^{l-1},..,2^{l-q},0\}$ where $q$ is the smallest integer above $log_2n^r=rlog_2n$.  Then $MAX[G',c] \geq 1/4MAX[G,c]$ since the effect of deleting all hyperedges with profits less than $2^{l-q}$ and also rounding down the remaining profits each independently reduce the maximum profit by a factor of at most $1/2$. Similarly let $c^*$ be the largest $c_i$ and $2^k$ the smallest power of $2$ at least $c^*$. Now group the vertices into buckets $V_i, i=1,..,s+3$ where $s$ is the smallest integer above $log_2n$. $V_i, i \leq s+2$ is the collection of vertices $v_j$ where $2^{k-i} < c_j \leq 2^{k+1-i}$ and $V_{s+3}$ is the collection of vertices $v_j$ where $0< c_j \leq 2^{k-s-2}$. It follows that the collection of smallest cost vertices $V_{s+3}$ has vertices with costs at most $B/(4n)$. \\ \\
{\em Three subgraph classes} \\
The hyperedge and vertex profits of $G'$ can be shared among at most $(rlog_2n+1)(log_2n+3)^r+1$ subhypergraphs at most three classes. These classes are defined by the cost groupings and profit roundings as follows: \\ \\
Class 1: The subhypergraph of $G'$ obtained by removing all of the edges (while retaining the vertex costs and profits).\\
Class 2: Subhypergraphs with edge profits all the same power of 2, and with vertex costs contained among at most $r$ of the cost buckets $\{V_i\}$, including $V_{s+3}$. These subhypergraphs have no vertex profits.\\
Class 3: Subhypergraphs with edge profits all the same power of 2, and with vertex costs contained among at most $r$ of the cost buckets $\{V_i\}$ not including $V_{s+3}$. These subhypergraphs have no vertex profits.\\
We shall show how to obtain $O(n^{\alpha_r+\epsilon})$ or better approximations for each of these classes in time $O(n^{5(r-2)-1+9/\epsilon})$ or less.\\ \\
Class 1: This class corresponds to the Knapsack Problem and so has an algorithm that has an approximation ratio of $1+\epsilon$ and a run time of $O(n^3/\epsilon)$ \cite{VAZ2003}, \cite{GAR1979}. \\ \\
Class 2: Any subhypergraph in this class takes the form of an $r$-partite hypergraph $H[A_1, A_2,..,A_r]$ with the vertices of each $A_i$ all in the same cost bucket. Also $A_1 \subseteq V_{s+3}$, with the costs of $A_j, j \geq 2$ between $a_j$ and $2a_j$, where $a_j$ is a positive power of 2, and $a_2 \leq a_3 \leq .. \leq a_r$. Let $H^{MAX}[B]=H[A'_1, A'_2,..,A'_r]$ where $A'_i \subseteq A_i$. We consider two cases. \\ 
Case 1 For some $i=2,3,..,r$ $|A'_i|<4$. In this case in finding a maximum solution it is sufficient to consider all choices of subsets $X$ of at most $3$ vertices from $A_i$ as part of a solution. The choice of the remaining vertices amounts to an instance of the SUK problem of order $r-1$. We have at most 7 such subsets $X$ and we can use the induction hypothesis to approximate the SUK problems of order $r-1$.\\ 
Case 2 $|A'_i| \geq 4$ for $i=2,3,..,r$. Now partition $A'_2$ into 4 subsets of as equal size as possible. Then for at least one of those subsets say $A''_2$ we have
\begin{equation}
|E[H[A'_1,A''_2,A'_3,..,A'_r]]| \geq \frac{1}{4}|E[H[A'_1,A'_2,A'_3,..,A'_r]]|.
\end{equation}
We may similarly construct subsets $A''_j$ of $A'_j, j=3,4,..,r$ so that
\begin{equation}
|E[H[A'_1,A''_2,A''_3,..,A''_r]]| \geq \frac{1}{4^{r-1}}|E[H[A'_1,A'_2,A'_3,..,A'_r]]|.
\end{equation}
By construction we have $cost(A''_i) \leq (4/7)cost(A'_i)$ for $i=2,3,..,r$ (this bound is attained with $A''_i$ containing two vertices of cost say $2x$, with the remaining 3 subsets in the partition containing one vertex each with cost $x$). Also $cost(A_1) \leq B/4$ since the vertices of $A_1$ have costs at most $B/(4n)$. It follows that 
$H[A_1, A''_2,..,A''_r]$ has a total vertex cost of $(4/7+1/4)B<B$ and so is feasible and
\begin{eqnarray}
|E[H[A_1,A''_2,A''_3,..,A''_r]]| &\geq& |E[H[A'_1,A''_2,A''_3,..,A''_r]]| \nonumber \\
&\geq& \frac{1}{4^{r-1}}|E[H[A'_1,A'_2,A'_3,..,A'_r]]|.
\end{eqnarray}
It follows that in Class 2 we can devise a strategy as follows. Corresponding to Case 1 select all combinations of 3 or less vertices from each $A_i, i=2,3,..,r$ in turn and by the induction hypothesis solve the SUK problem of order $r-1$ corresponding to the choice of the remaining vertices. This process takes a total time $O(((r-1)n^3)n^{5(r-3)+9/\epsilon})$. Since $r-1<n$ this is at most $O(n^{5(r-2)-1+9/\epsilon})$. This process also produces an approximation ratio $O(n^{\alpha_{r-1}+\epsilon})$. Now corresponding to Case 2 assume that all of $A_1$ is part of a solution and choose the remaining vertices again as an instance of the SUK problem of order $r-1$. This takes time $O(n^{5(r-3)+9/\epsilon})$ and by equation (21) has an approximation ratio of within $O(4^{r-1}n^{\alpha_{r-1}+\epsilon})=O(n^{\alpha_{r-1}+log4/logn+\epsilon})$. Thus if the maximum profit found among all the solutions found in Cases 1 and 2 is retained it takes at most time $O(n^{5(r-2)-1+9/\epsilon})$ with approximation ratio at most $O(n^{\alpha_{r-1}+log4/logn+\epsilon})$. Clearly this approximation ratio is less than $O(n^{\alpha_{r}+\epsilon})$\\ \\
Class 3: Any subhypergraph in this class takes the form of an $r$-partite hypergraph $H[A_1, A_2,..,A_r]$ with the vertices of each $A_i$ all in the same cost bucket, with the costs non-decreasing with $i$. We can simplify the analysis by scaling the profits and costs. Specifically let the costs of the vertices in $A_1$ be between $\lambda$ and $2\lambda$ and the hyperedge profits all $\kappa$. Form a modified subhypergraph $\tilde{H}$ from $H$ by dividing the costs by $\lambda$, and setting the edge profits to $1$. Also set the cost limit $\tilde{B}$ for $\tilde{H}$ to $\tilde{B}=B/\kappa$. Then in $\tilde{H}$ the costs of $A_1$ are between $1$ and $2$, and the hyperedge profits are all $1$. Also
\begin{equation}
MAX[H,B]=\frac{1}{\kappa}MAX[\tilde{H},\tilde{B}].
\end{equation}
Thus approximating SUKP for $H,B$ is equivalent to approximating the SUKP for $\tilde{H},\tilde{B}$. For this reason we may assume without loss of generality that the hypergraph $H[A_1, A_2,..,A_r]$ is scaled with costs of $A_1$ between $1$ and $2$, and hyperedge profits all $1$.

Now let $H^{MAX}[B]=H[A'_1, A'_2,..,A'_r]$ where $A'_i \subseteq A_i$. Let $\gamma$ be given by
\begin{equation}
\gamma=1+\alpha_{r-1}-\alpha_r.
\end{equation}
We take a number of cases.\\
Case 1 $a_r \geq n^\gamma$. We take two further subcases. \\
Case 1.1 $|A'_r|\leq 6n^{1-\gamma}$. In this case there must be some vertex $x$ of $A_r$ for which 
\begin{equation}
MAX[H[A_1, A_2,..,A_{r-1},x],B] \geq \frac{1}{6n^{1-\gamma}}MAX[H[A_1, A_2,..,A_r],B]
\end{equation}
The SUK problem for $H[A_1, A_2,..,A_{r-1},x]$ can now be solved by forming the hypergraph $H^x$ of order $r-1$ by deleting $x$ from each hyperedge of $H[A_1, A_2,..,A_{r-1},x]$. Then
\begin{equation}
MAX[H^x,B-c_x]=MAX[H[A_1, A_2,..,A_{r-1},x],B].
\end{equation}
By our induction hypothesis we can approximate $MAX[H^x,B-c_x]$ to at most a factor of $n^{\alpha_{r-1}+\epsilon}$ and so approximate $MAX[H[A_1, A_2,..,A_r],B]$ to at most a factor of $6n^{1-\gamma+\alpha_{r-1}+\epsilon}=6n^{\alpha_r+\epsilon}$. This takes time at most $O(nn^{5(r-3)+9/\epsilon})=O(n^{5(r-3)+1+9/\epsilon})$.\\ 
Case 1.2 $|A'_r| > 6n^{1-\gamma}$. We take two further subcases. \\ 
Case 1.2.1 $|A'_r|=1$ In this case we may consider the subgraphs of $H$ formed by selecting just one vertex from $A_r$. As above this leads to at most $n$ SUK problems of order $r-1$. \\ 
Case 1.2.2 $|A'_r| \geq 2$. In this case some half, or as close to half as possible of the vertices of $A'_r$, say $A''_r$ can be selected with
\begin{equation}
MAX[H[A_1, A_2,..,A''_r],B] \geq \frac{1}{3}MAX[H[A_1, A_2,..,A_r],B].
\end{equation}
Since the total cost of vertices in $A_r$ is at least $6n$ it follows that if we assume that all the vertices of $A_1$ 
are used in a vertex selection then some selection of the remaining vertices must produce an optimum at most at least $1/3$ of $MAX[H[A_1, A_2,..,A_r],B]$. This assumption however leads to an SUK problem of order $r-1$ which can be solved by the induction hypothesis. \\ 
Case 2 $a_r < n^\gamma$. We consider two subcases. \\ 
Case 2.1 $B \leq 2ra_r$. In this case $|A'_r| \leq 2r$ and we can solve as in Case 1 of Class 2. \\ 
Case 2.2 $B > 2ra_r$. This case is the most challenging and we provide a graph transformation method to deal with it. This method is the hypergraph equivalent to the method given in \cite{TAY22015} (see Section 2.3, Class 4, Case 2). In the following the inequality defining this case will allow us to consider fractions such as $B/(2ra_j) > 1$ as if they were integers without affecting the thrust of the argument.

Construct a graph $H^*$ from $H$ by replacing each $A_j$ by $a_j$ copies $A_j^1,..,A_j^s$ of $A_j$ where $s=a_j$. Each hyperedge $e$ of $H$ corresponds to $\prod_{j}a_j$ hyperedges of $H^*$ specified as follows. Let $A_j=\{x_{jl}:l=1,..,a_j\}, A_j^p=\{x_{jl}^p:l=1,..,a_j\}, j=1,..,r, p=1,..,a_j$. Then $\{x_{1i},x_{2j},..,x_{rz}\}$ is a hyperedge of $H$ iff $\{x_{1i}^s,x_{2j}^t,..,x_{rz}^u\}$ are hyperedges of $H^*$ for any $s=1,..,a_1, t=1,..,a_2,...,u=1,..,a_r$. In other words for any $r$ positive integers $i \leq a_2, j \leq a_3, ...,z \leq a_r$,
\begin{equation}
H^*[A_1^1, A_2^i,A_3^j,..,A_r^z] \cong H[A_1, A_2,A_3,..,A_r].
\end{equation}
In $H^*$ set $cost(x_{jl}^p)=(1/a_p)cost(x_{jl})$ and so the costs of vertices of $H^*$ are between 1 and 2. The transformation from $H$ to $H^*$ is illustrated in figure 2. 
\begin{figure}
\begin{center}
\includegraphics[width=100mm]{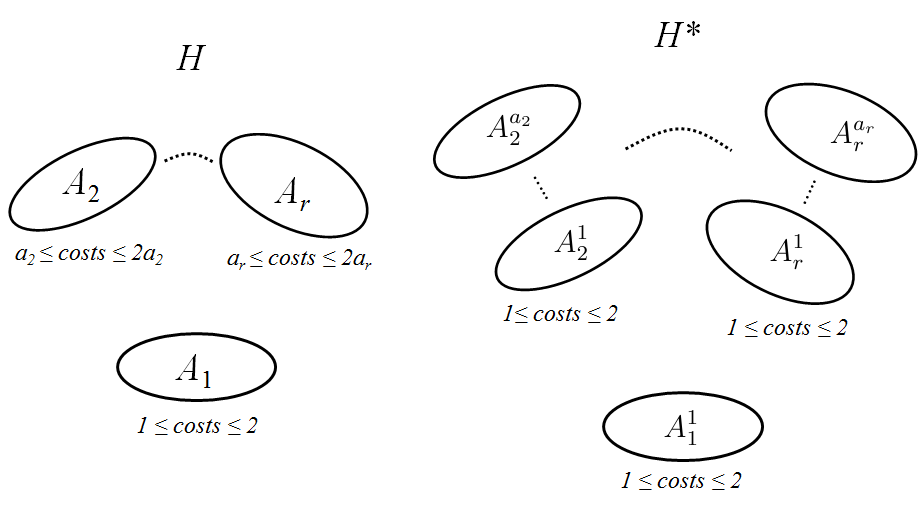}
\caption{\label{figure1.png} hypergraphs $H$ and $H^*$}
\end{center}
\end{figure}
Now 
\begin{equation}
MAX[H^*,B] \geq (\prod_{j}a_j)MAX[H,B]
\end{equation}
since any feasible subgraph of $H$ can be replicated $\prod_{j}a_j$ times to a feasible solution of $H^*$. We approximate the DkSH problem for $H^*$ with $k=B$ to at most an approximation ratio of say $q$. Since $H^*$ has at most $a_rn$ vertices, and this is at most $n^{1+\gamma}$ by the case 2 clause, then $q=O(n^{(1+\gamma)\theta_r+\epsilon})$ where $\theta_r=r/2-1/2-1/(2r)$ (see Theorem 1.1). We note that it is elementary to verify that 
\begin{equation}
(1+\gamma)\theta_r=(2+\alpha_{r-1}-\alpha_r)\theta_r=\alpha_r.
\end{equation}
Thus $q=O(n^{\alpha_r+\epsilon})$. For ease of notation we aggregate the vertices of $H^*$ into sets $C_j, j=1,..,r$ where $C_j=\bigcup_pA_j^p$. Let $H^{**}=H^*[C'_1,C'_2,...,C'_r]$ with $C'_j \subseteq C_j$ be an induced subgraph of $H^*$ corresponding to this approximation. Thus
\begin{equation}
|E[H^{**}]| \geq \frac{1}{q}MAX[H^*,B].
\end{equation}
The time taken for the DkSH calculation above is at most
\begin{equation}
O([n^{1+\gamma}]^{r-2+1/\epsilon}).
\end{equation}
Since $1+\gamma < 2$ this time factor is at most $O(n^{2r-4+2/\epsilon})$. 
By stages we construct a subgraph of $H^{**}$ that gives us the required approximation to $MAX[H,B]$. Let $\delta_i$ to be the degree of the vertex $x_{1i}^1$ in $H^{**}$. Select $D_1$ to be any selection of vertices $x_{1i}^1$ with the highest $B/2r$ degrees among $\{\delta_i\}$. Then since $C'_1$ contains at most $B$ vertices
\begin{equation}
eH^*[D_1,C'_2,...,C'_r] \geq \frac{1}{2r}eH^*[C'_1,C'_2,...,C'_r].
\end{equation}
Now define $\delta_{ij}$ to be the degree of $x_{2i}^j$ in $H^*[D_1,C'_2,...,C'_r]$ and $\delta_i^*=max_j\{\delta_{ij}\}$. Now choose a subset $D_2$ of $A_2^1$ to be those $x_{2i}^1$ for which $\delta_i^*$ is among the $B/(2ra_2)$ highest values among $\{\delta_i^*\}$. Then since $C'_2$ contains at most $B$ vertices
\begin{eqnarray}
|E[H^*[D_1,D_2,C'_3...,C'_r]| &\geq& \frac{1}{2ra_2}|E[H^*[D_1,C'_2,...,C'_r]]| \nonumber \\
&\geq& \frac{1}{(2r)^2a_2}|E[H^*[C'_1,C'_2,...,C'_r]]|.
\end{eqnarray}
We continue in this way to select $B/(2ra_j)$ vertices $D_j$ of the form $x_{ji}^1$ from $C'_j, j=3,..,r$ so that
\begin{equation}
|E[H^*[D_1,D_2,...,D_r]]| \geq \frac{1}{(2r)^r\prod a_j}|E[H^*[C'_1,C'_2,...,C'_r]]|.
\end{equation}
Noting that $|E[H^*[C'_1,C'_2,...,C'_r]]| \geq 1/qMAX[H^*,B]$ by combining inequalities (35) and (29)
\begin{equation}
|E[H^*[D_1,D_2,...,D_r]]| \geq \frac{1}{q(2r)^r\prod a_j}MAX[H^*,B]  \geq \frac{1}{q(2r)^r}MAX[H,B].
\end{equation}
Since $H^*[D_1,D_2,...,D_r]$ is trivially isomorphic to a subgraph of $H$, we have constructed an order $q$ approximation method for $MAX[H,B]$. This completes Case 2.2 and Class 3. In summary for all the cases in Classes 1, 2, and 3 approximations to $MAX[H,B]$ are found of order at most $O(n^{\alpha_r+\epsilon})$ in a time at most order $O(n^{5(r-2)-1+9/\epsilon})$.

To complete the algorithm we generate all $O((rlog_2n+1)(log_2n+3)^r+1)$ of the subhypergraphs $H$ and find the largest profit among the approximations to $MAX[H,B]$. Then $G'^{MAX}[B]$ must share a vertex/edge profit of at least 
\begin{equation}
\frac{MAX[G',B]}{(rlog_2n+1)(log_2n+3)^r+1}
\end{equation}
with at least one of the subgraphs $H$. Thus the largest approximation found to $MAX[H,B]$ over the subhypergraphs $H$ must approximate $MAX[G',B]$ to within a factor of at most
\begin{equation}
O([(rlog_2n+1)(log_2n+3)^r+1]n^{\alpha_r+\epsilon}).
\end{equation}
Finally since $MAX[G',B] \geq 1/4 MAX[G,B]$ this subhypergraph must approximate $MAX[G,B]$ to within a factor of at most
\begin{equation}
O(4[(rlog_2n+1)(log_2n+3)^r+1]n^{\alpha_r+\epsilon}).
\end{equation}
This approximation is in turn less than $O(n^{\alpha_r+2\epsilon})$ for large $n$. The time taken is at most of order
\begin{equation}
O([(rlog_2n+1)(log_2n+3)^r+1]n^{5(r-2)-1+9/\epsilon})
\end{equation}
which is similarly less than $O(n^{5(r-2)+9/\epsilon})$ for large $n$. This completes the induction step and so the proof.

In Table 1 we provide the values of the indices $\theta_m$ and $\alpha_m$ for $m=2,...,6$ corresponding to the approximation ratios $O(n^{\theta_r+2\epsilon})$ and $O(n^{\alpha_r+\epsilon})$ for the DkSHP and SUKP respectively.

\begin{table}
\caption{Approximation Ratio Indices for DkSHP and SUKP}
\centering
\begin{tabular}{c c c c c c}
\hline\hline
$m$ & $2$ & $3$ & $4$ & $5$ & $6$ \\
\hline
$\theta_m$ & 1/4 & 5/6 & 11/8 & 19/10 & 29/12 \\
$\alpha_m$ & 2/5 & 12/11 & 34/19 & 72/29 &  130/41\\

\hline
\end{tabular}
\label{table:nonlin}
\end{table}







\end{document}